\documentclass[20pt]{article}

\usepackage{PRIMEarxiv}

\usepackage[utf8]{inputenc} 
\usepackage[T1]{fontenc}    
\usepackage{hyperref}       
\usepackage{url}            
\usepackage{booktabs}       
\usepackage{amsfonts}       
\usepackage{nicefrac}       
\usepackage{microtype}      
\usepackage{lipsum}
\usepackage{fancyhdr}       
\usepackage{graphicx}       
\graphicspath{{media/}}     
\usepackage{subcaption}

\usepackage[utf8]{inputenc}
\usepackage[T1]{fontenc}
\usepackage{dirtytalk}
\usepackage{hyperref}
\hypersetup{
    colorlinks=false,
    linkcolor=blue,
    filecolor=blue,      
    urlcolor=blue,
    citecolor = black,
}
\title{Automatic Detection of COVID-19 from Chest X-ray Images Using Deep Learning Model
\thanks{\textit{\underline{Citation}}: 
\textbf{Das, A., Agarwal, R., Singh, R., Chowdhury, A.,  Nandi, D. (2022, March). Automatic detection of COVID-19 from chest x-ray images using deep learning model. In AIP Conference Proceedings (Vol. 2424, No. 1). AIP Publishing.}}
}

\author{
  Alloy Das, Rohit Agarwal, Rituparna Singh, Arindam Chowdhury \\
  University Institute of Technology\\
  Burdwan University\\
  West Bengal, India \\
  \texttt{\{alloyuit, rohit8.1996, singhri085\}@gmail.com, arindamchowdhury@uit.buruniv.ac.in} \\
   \And
  Debashis Nandi \\
  National Institute of Technology\\
  Durgapur\\
  West Bengal, India\\
  \texttt{debashisn2@gmail.com} \\
}

\begin{document}
\maketitle

\begin{abstract}
The infectious disease caused by novel corona virus (2019-nCoV) has been widely spreading since last year and has shaken the entire world.
It has caused an unprecedented effect on daily life, global economy and public health. 
Hence this disease detection has life-saving importance for both patients as well as doctors.
Due to limited test kits, it is also a daunting task to test every patient with severe respiratory problems using conventional techniques (RT-PCR) .
Thus implementing an automatic diagnosis system is urgently required to overcome the scarcity problem of Covid-19 test kits at hospital, health care systems. 
The diagnostic approach is mainly classified into two categories-laboratory based and Chest radiography approach. 
In this paper, a novel approach for computerized corona virus (2019-nCoV) detection from lung x-ray 
images is presented. Here, we propose models using deep learning to show the effectiveness of diagnostic
systems. In the experimental result, we evaluate proposed models on publicly available data-set which exhibit satisfactory performance and promising results compared with other previous existing methods.
\end{abstract}

\keywords{Deep Learning \and Chest X-Ray \and nCOVID-19 Detection }

\section{Introduction}
The deadlier infection caused by corona virus disease (SARS-CoV-2) was first discovered in Wuhan, Hubei
Province, China since December 2019.  This virus is transmitted through respiratory droplets, contact 
and soon it rapidly spreads to other domestic cities and countries beyond China and has swept the entire
world. Covid-19 is often deadlier in people 60+ years or people with co-morbidity such as lung or heart
diseases, diabetes or conditions that affect  immune system. Now, it is taking place in all ages and if
getting seviour it is causing death. Symptoms show up in people within two to 14 days due to exposure to the virus. 

During this pandemic, countless people were affected and several people died. But number of Test kits is not sufficient to detect the disease and experts in this domain are also limited. Hence it necessitates an alternative automated diagnostic system that may provide secondary opinion to health care experts to identify infected persons from healthy one.

Chest X-ray plays a vital role since it acts as a non-invasive clinical adjunct possessing discriminative features for identifying pulmonary diseases. Almost half of patients with COVID-19 positive symptoms have significant abnormality in chest x-ray with peripheral GGO(Ground Graph Opacification) affecting the lower lobes being the most common one. ``If people have a mild illness(asymptomatic) or doubt they should go for a chest x-ray''  as suggested by (AIIMS) Director at a press conference. Chest x-ray can be used in detecting and diagnosis of Covid-19 affected patients.  Thus it may act a secondary screening tool for the detection of 2019-nCoV where expert radiologists can identify infectious lesions associated with Covid-19. Fig 1 shows normal and infected people with mark able visual characteristics in chest X-ray.
\begin{figure}[h]
\begin{subfigure}{.3\textwidth}
  \centering
  \includegraphics[width=.8\linewidth]{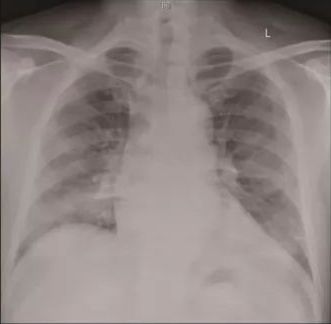}  
  \caption{}
  \label{fig:sub-third}
\end{subfigure}
\begin{subfigure}{.3\textwidth}
  \centering
  \includegraphics[width=.8\linewidth]{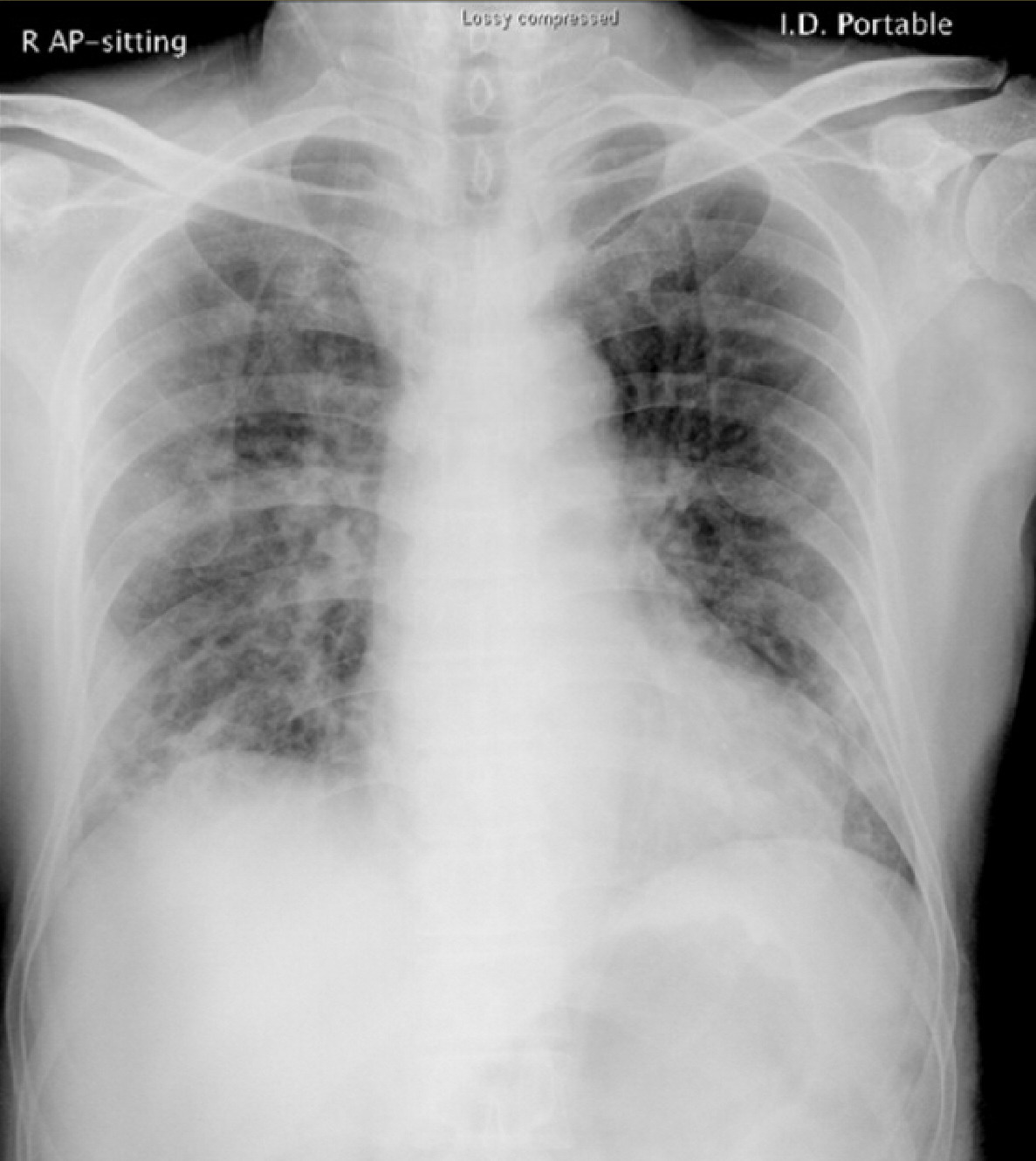}  
  \caption{}
  \label{fig:sub-fourth}
\end{subfigure}
\begin{subfigure}{.3\textwidth}
  \centering
  \includegraphics[width=.8\linewidth]{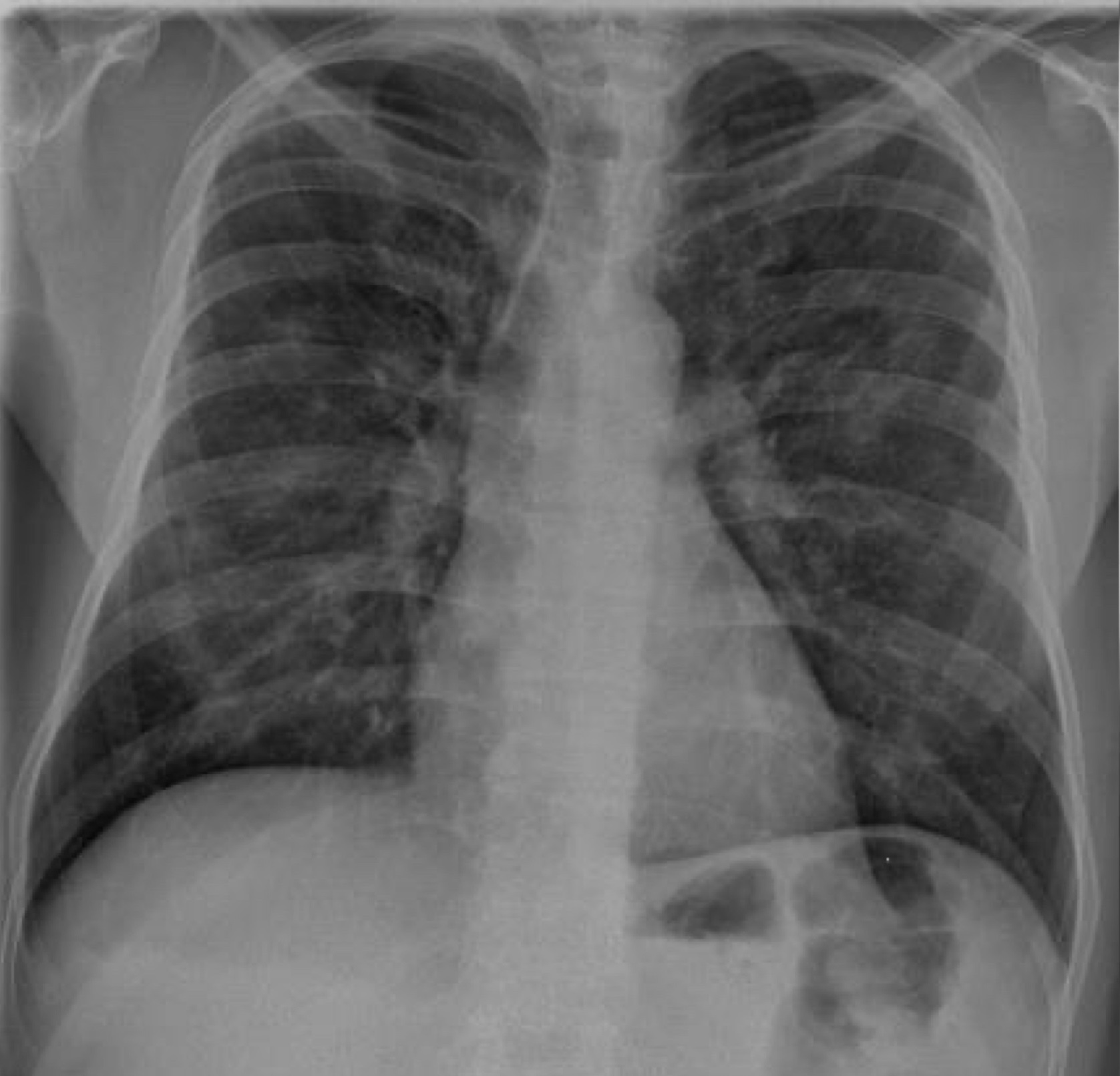}  
  \caption{}
  \label{fig:sub-third}
\end{subfigure}
\begin{subfigure}{.3\textwidth}
  \centering
  \includegraphics[width=.8\linewidth]{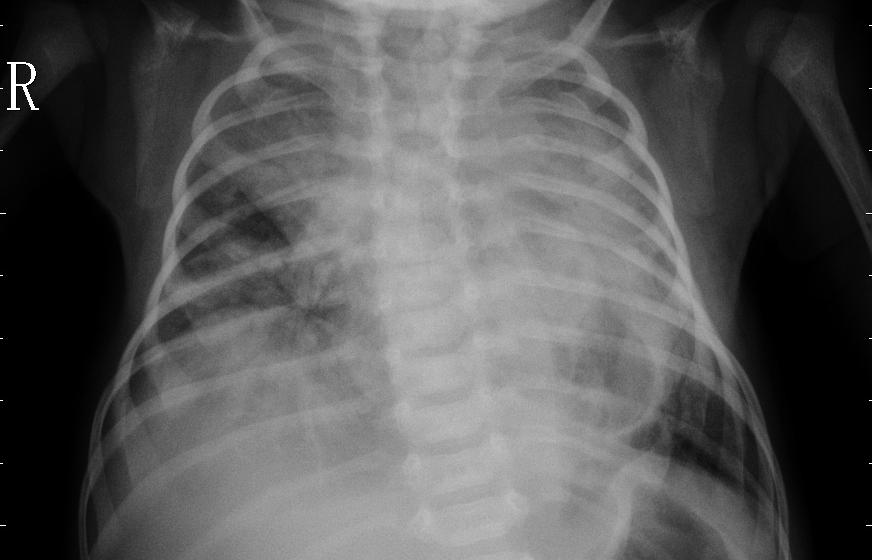} 
  \caption{}
  \label{fig:sub-third}
\end{subfigure}
\begin{subfigure}{.3\textwidth}
  \centering
  \includegraphics[width=.8\linewidth]{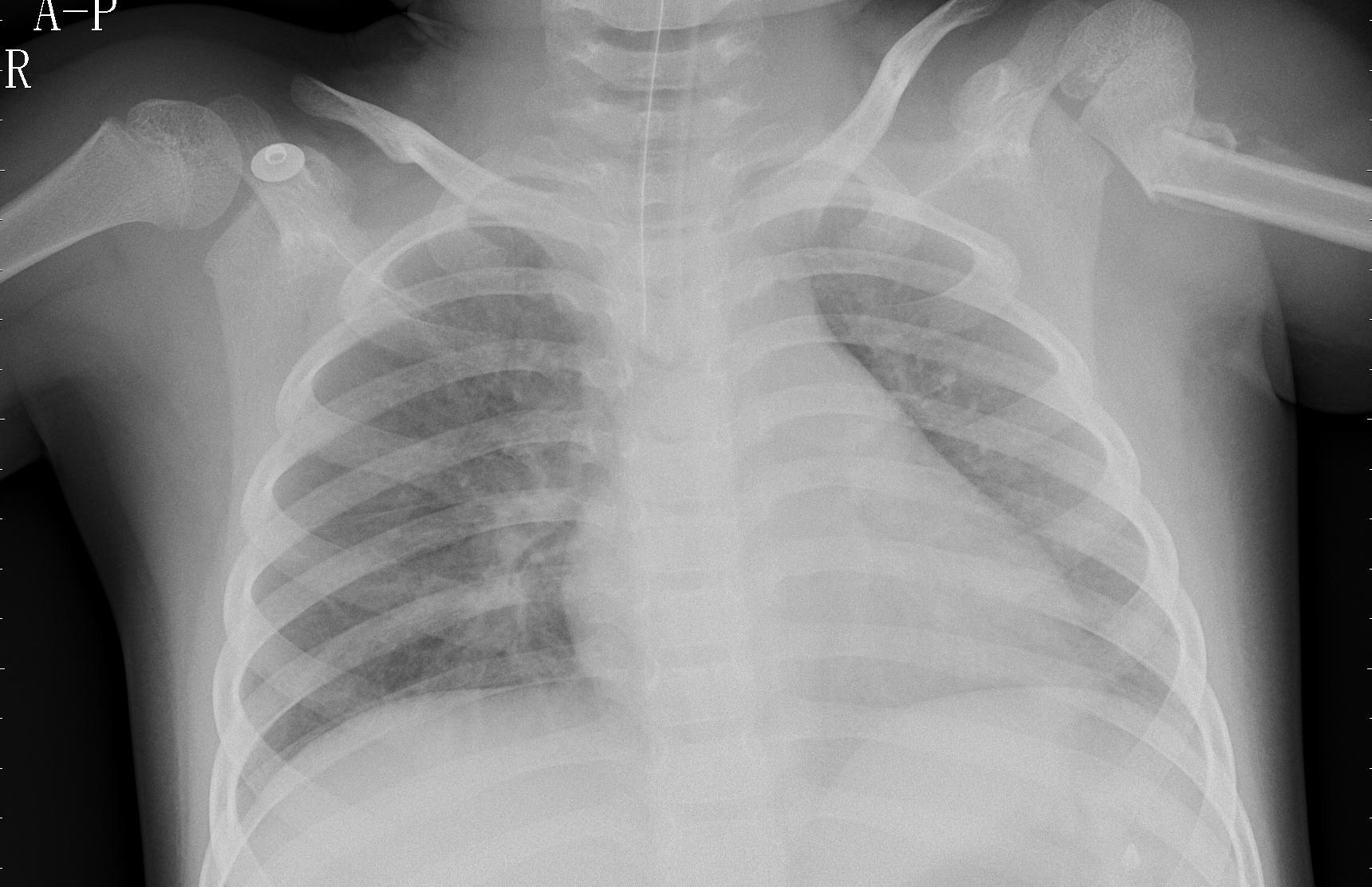} 
  \caption{}
  \label{fig:sub-third}
\end{subfigure}
\begin{subfigure}{.3\textwidth}
  \centering
  \includegraphics[width=.8\linewidth]{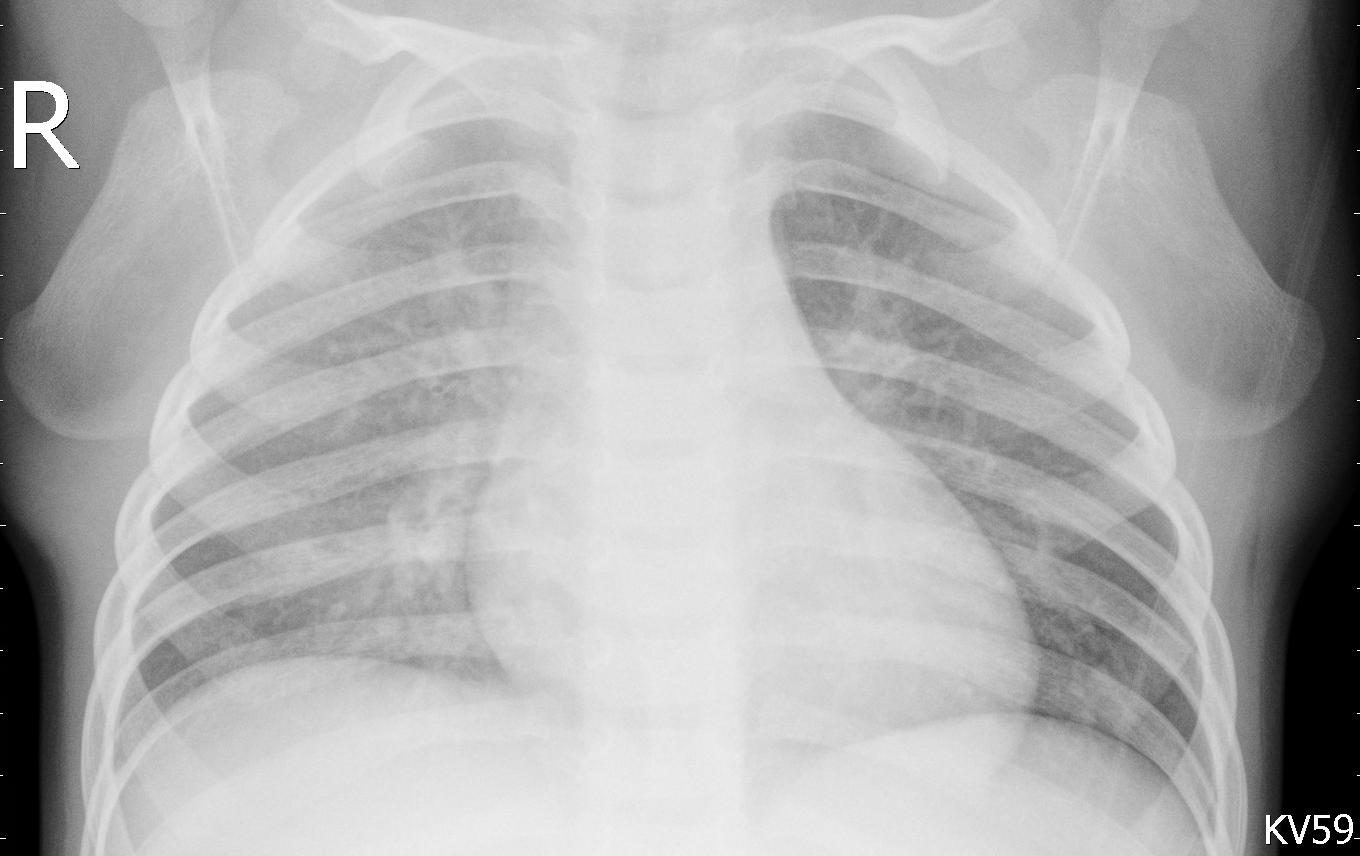} 
  \caption{}
  \label{fig:sub-third}
\end{subfigure}
\begin{subfigure}{.3\textwidth}
  \centering
  \includegraphics[width=.8\linewidth]{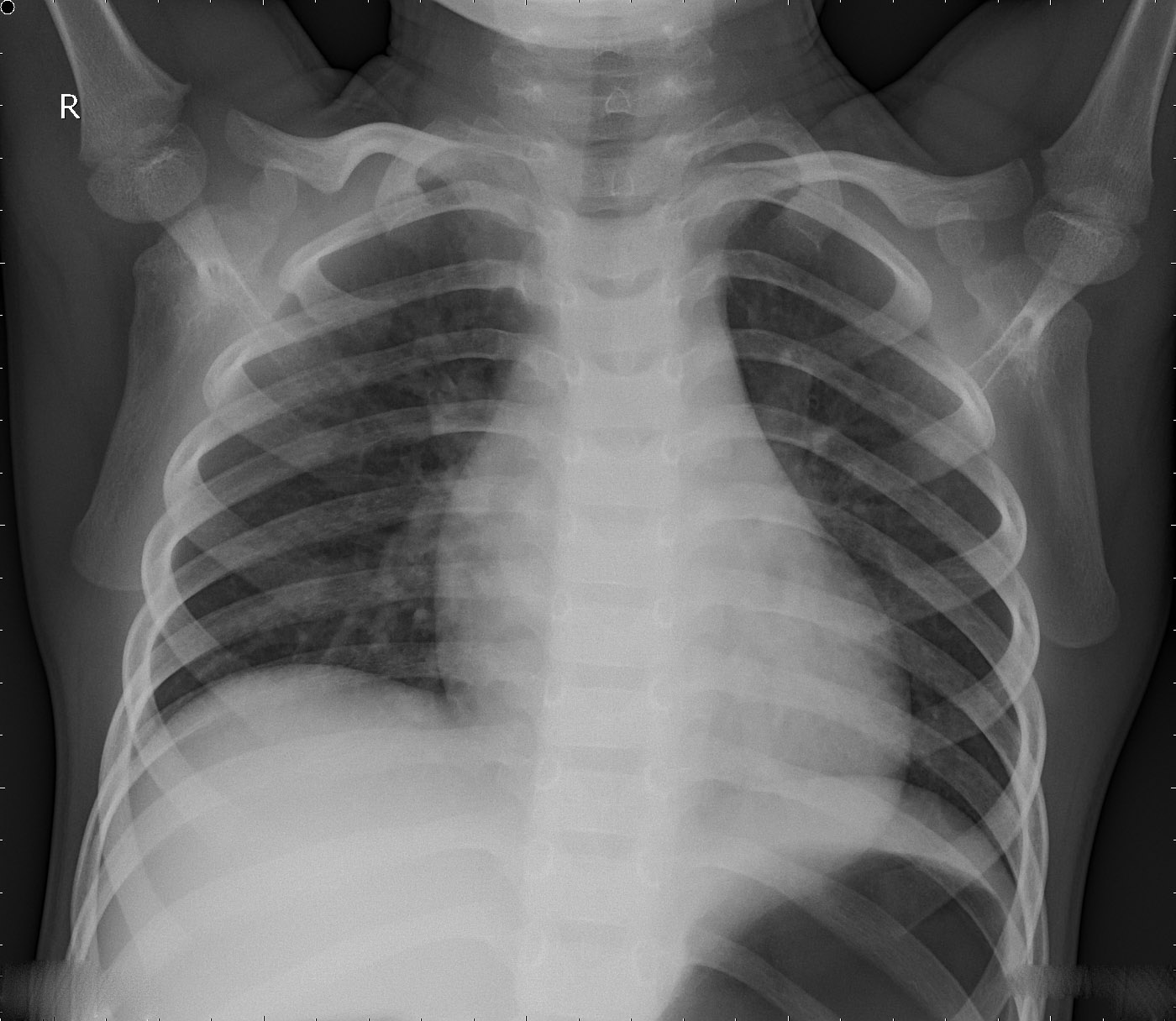}  
  \caption{}
  \label{fig:sub-first}
\end{subfigure}
\begin{subfigure}{.3\textwidth}
  \centering
  \includegraphics[width=.8\linewidth]{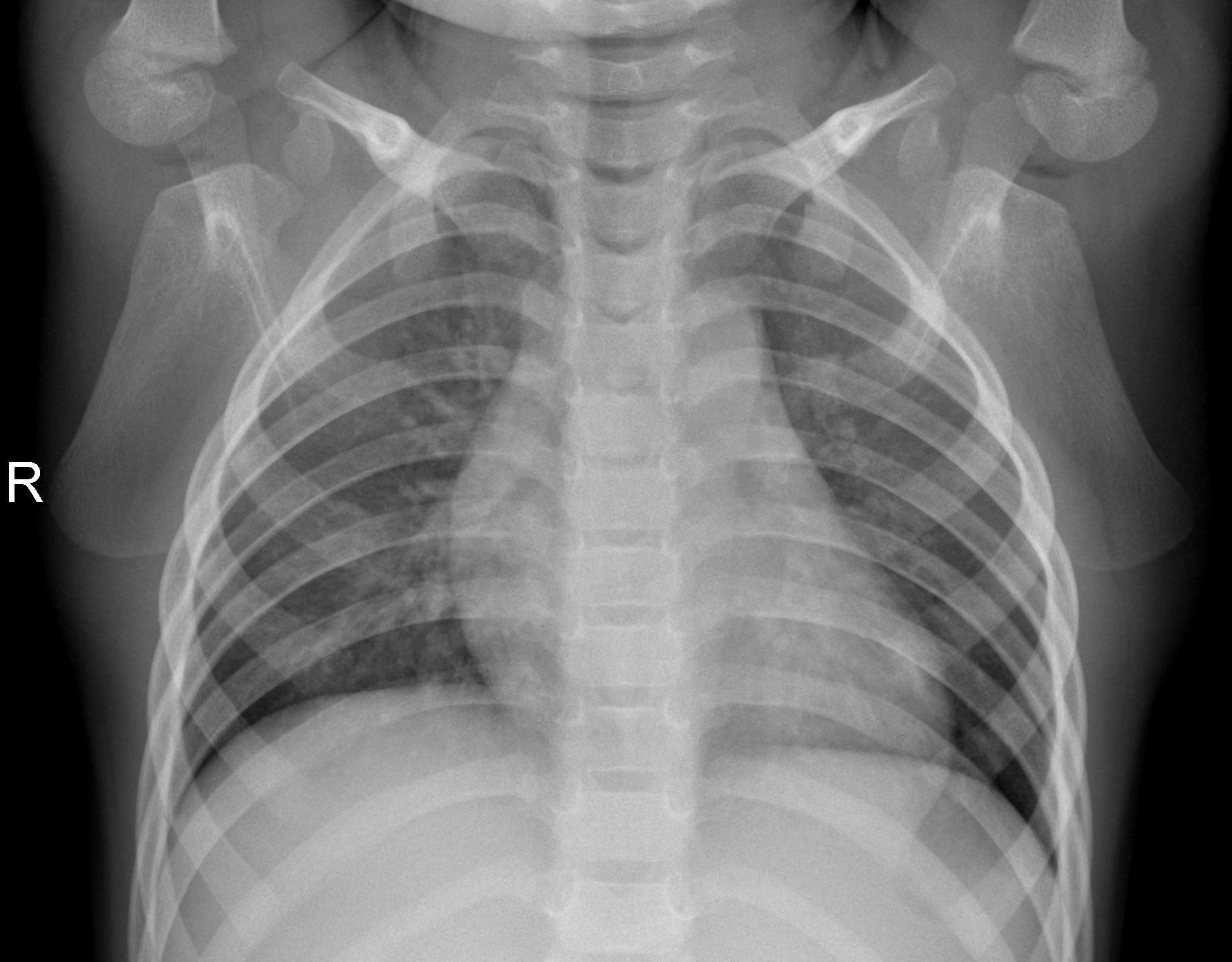}  
  \caption{}
  \label{fig:sub-first}
\end{subfigure}
\begin{subfigure}{.3\textwidth}
  \centering
  \includegraphics[width=.8\linewidth]{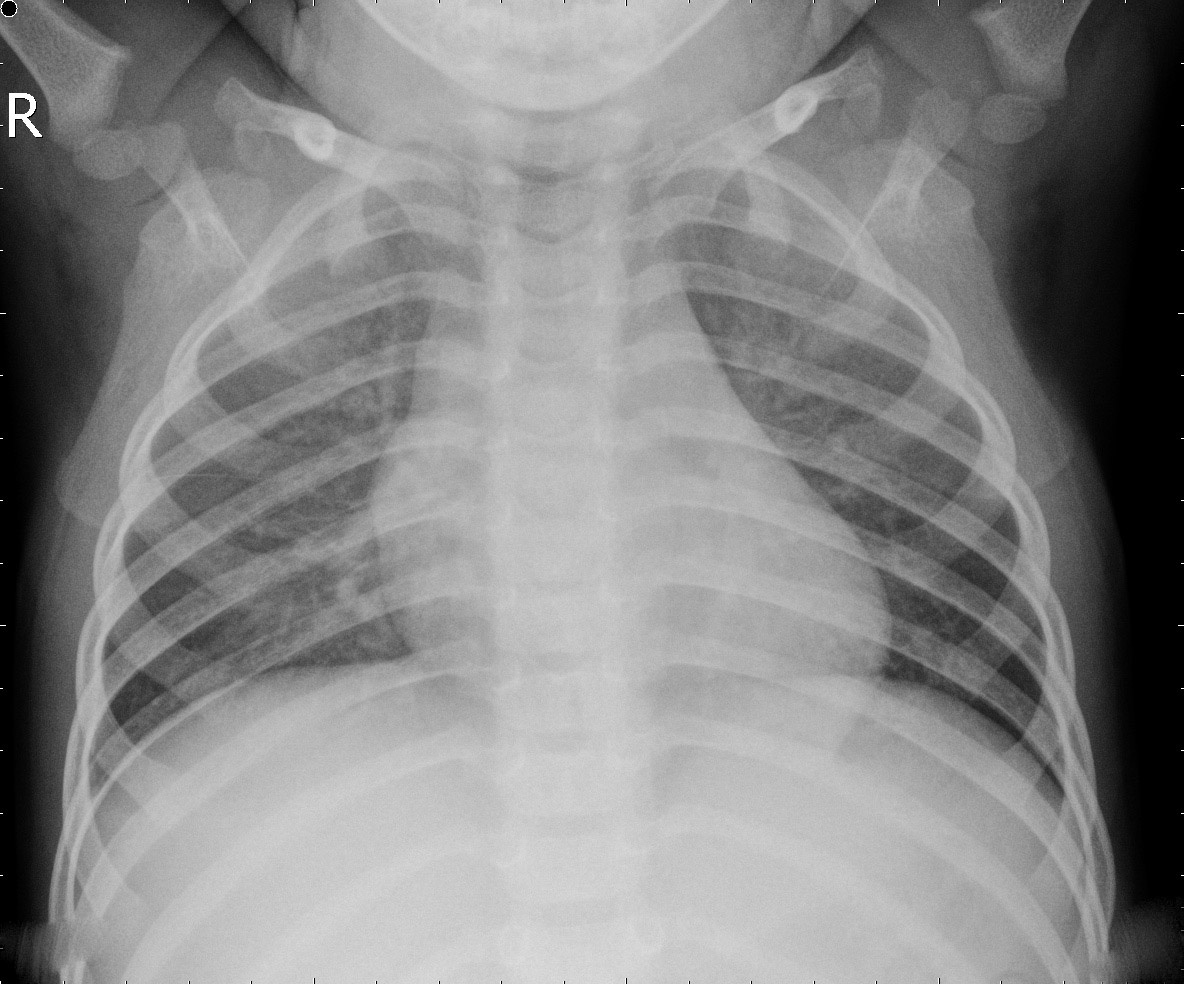}  
  \caption{}
  \label{fig:fig}
\end{subfigure}
\caption{(a–c) 2019-nCoV affected chest X-Ray images (d–f) Pneumonia affected chest X-Ray (g-i) Normal chest X-Ray images.}
\label{fig:fig}
\end{figure}

\section{Literature Review}

To fight the nCOVID-19 pandemic, the recent advancement in artificial intelligence and machine learning is used to develop an automated computer-aided diagnostic system. Deep Learning specialized research area in AI and machine learning, enables to create end-to-end models and produce promising results with input data but it bypasses feature extractions(manual) procedure. Several deep learning based models were proposed by researchers.

Arpan et al.\cite{a} designed a deep neural network based detector and got $90.5$\% accuracy with $100$\% sensitivity tested on publicly available covid chest x-ray data-set. Deep learning model based on ResNet-$101$ architecture was performed by Azemin et al\cite{b}. CNN architecture was successfully used to detect covid-$19$ disease from chest X-ray images and got $98.50$\% accuracy\cite{c}. Stephanie et al. performed a survey based study to determine the sensitivity of chest x-ray image towards covid-19 detection\cite{d}. DarkNet model with You Only Look Once(YOLO) object detection was used by Ozturk et al\cite{e} produced 98.08\% classification accuracy for two classes and  87.02\%  accuracy for three classes. An ensemble based voting method applied by Chandra et al.\cite{f}. Ahmed et al.\cite{g} proposed a  architecture HRNet  used mainly for features extraction and embedding this approach with the U-net model. L. Wang et al. (2020)\cite{h} proposed a network COVID-Net for the detection of Covid-19 cases and produced three-class classification accuracy of 93.00. Abbas, Abdelsamea, and Gaber (2020)\cite{i} uses a deep CNN model called DeTrac to identify the diseases from chest x-ray images with 95.12\% classification accuracy for three classes.

Our proposed model produces precise diagnostics for two class classification (Covid vs. Normal) and three class classification (Covid vs. Normal vs. Pneumonia) on the publicly available covid-chestxray-dataset\cite{j} and produces a classification accuracy of $99.17$\%  with $100$\% sensitivity for two class classification using W-net model and 97.50\% with 100\% sensitivity for three class classification. Our method can be used to assist the radiologists in validating the initial screening results.

\section{Proposed Methodology}
\subsection{Network Architecture}

\hspace{1cm}The working principle of our proposed architecture for nCovid-19 detection is shown in Figure: 2
\begin{figure}[h]
\centering
    \includegraphics[scale=0.45]{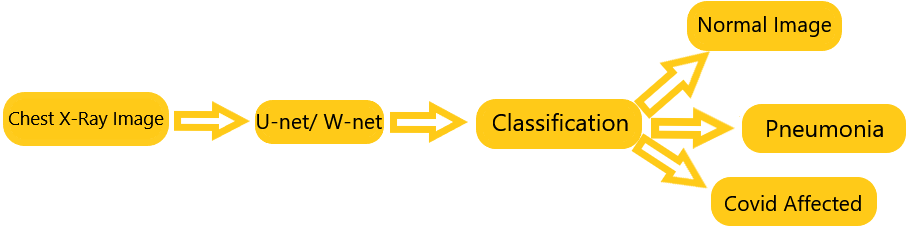}
    \caption{Workflow of the proposed classification system }
    \label{fig:myfigure}
\end{figure}

In order to do semantically segmentation on chest x-ray images, a well-known network U-Net is used\cite{k}. U-net has down-sampling and the up-sampling part where consecutive blocks of convolution and max-pooling block are used and it keeps on repeating the same process on both encoder and decoder part. It also transfers a layer of features from the encoding part and append it to the decoder part.

\subsection{W-Net}

\hspace{1cm} We have used one convolution network W-net for image segmentation. We have taken input images having a size of $400\times400$ pixels and all are gray-scale images in one single channel. Then we run a convolution of $3\times3$ with the transfer function ReLU. At this time, we have 32 channels or kernels. After two consecutive convolutions, we have a max-pooling of size $2\times2$. So, now the kernel size is 64 but in the X and Y space image size is reduced down by 2. Now the size of images is $200\times200$. Then we have consecutive two convolutions $3\times3$ with the transfer function ReLU. After that, again max-pooling of $2\times2$ is done. So, the size of images is reduced down by 2 and the size of images is $100\times100$. Now we have 128 kernel sizes and image sizes $100\times100$. After two consecutive convulsions, we have max-pooling.
\begin{figure}[ht]
    \includegraphics[scale=0.3]{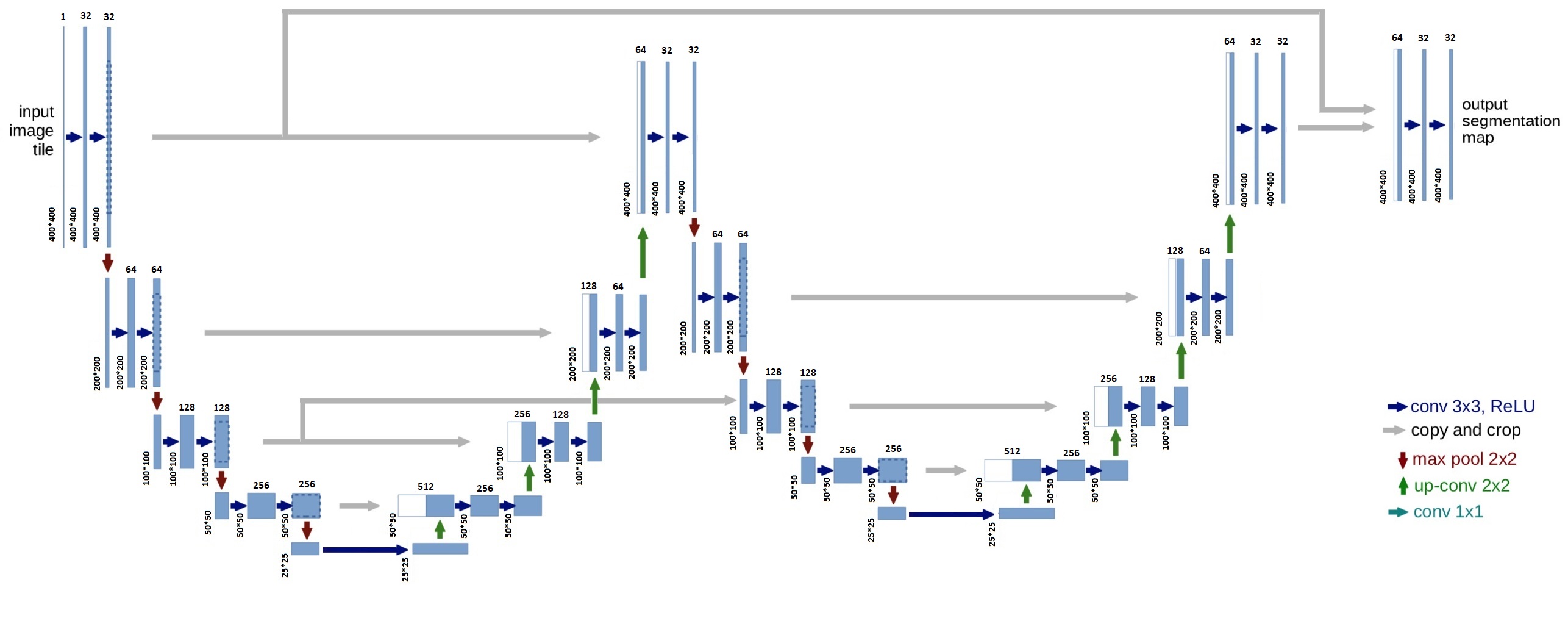}
    \caption{W-NET}
    \label{fig:myfigure}
\end{figure}
This reduces the image size and the kernel size is 256. Again there are two banks of convolutions and a max-pooling is applied. So, this is contacting path, next it starts expanding path. Now when we go up, we are applying up-sampling. So here we have image size $25\times25$ and after up-sampling it comes out $50\times50$. Now from the previous path, 256 channels are concatenated in it and new channel size is 512, here we have done copy and crop arrangements. Now, we are performing two consecutive convolutions  and a up-sampling operation. This process continues and reach to the upper layer. Then, we have $400\times400$ image size and 32 channels and we are performing two convolution operation. From here we go again for max-pooling $2\times2$ that reduces image size and the kernel size becomes 64. Then consecutively two convolutions and max-polling is done. From here we repeat the previous steps and lastly reach the upper level of expanding path. The sizes of images obtained are $400\times400$ and the kernel size is 32. In this way we perform down-sampling and up-sampling operations which gives us the output segmentation map.

\section{Experimental Result:}
Our framework was implemented under the open-source deep learning library Tensor-Flow\cite{l} on a server with Intel(R) Xeon(R) E5-2620 V3 2.30GHz CPU, 12GB NVIDIA Tesla K80 GPU and elementary OS 5.1.7 Hera as OS.
\subsection{Data-Set:}
We take CT scan images from publicly available  \href{run: https://github.com/ieee8023/covid-chestxray-dataset}{data-set}\cite{j}.
This data-set consists of $950$ images. The data-set is labeled in $25$ labels including viral, bacterial, fungal and lipoid pneumonia. Among them 196 images are labeled as Covid-19. We took all of the images. For normal images we use standard \href{run: https://data.mendeley.com/datasets/rscbjbr9sj/2}{data-set}\cite{m}. It consists of test train separated data along with Normal and Pneumonia. We take $400$ normal images from this data-set. Now our data-set consists of $596$ images for two class classification.

\hspace{1cm}After that we have added more $400$ images labeled as pneumonia from standard \href{run: https://data.mendeley.com/datasets/rscbjbr9sj/2}{data-set}  for three class classification. Now our new data-set consist of total $996$ images.  

\subsection{TRAIN-TEST SPLIT:}
\subsubsection{Two-Class Classification:}
For training and testing data we split the data set into $80:20$ ratio. In the training dataset there are $476$ images. The 1:3 ratio in labels are used here. So in training data there are $130$ COVID-$19$ images and $346$ Normal images. In the testing data set there are $39$ COVID-$19$ images and $81$ Normal images. Here we use two deep learning models U-Net and W-Net to perform on the newly created data-set.

\subsubsection{Three-Class Cassification:}
For training and testing data we split the data set into $80:20$ ratio. In the training dataset there are $796$ images. The 1:3:3 ratio in labels are used here. So in training data there are $156$ images as COVID-$19$ images,  $320$ images as Normal images and 320 images as Pneumonia images . In the testing data set there are $40$ COVID-$19$ images, $80$ Normal images and $80$ Pneumonia images. Here also we use two deep learning models U-Net and W-Net to perform on the newly created data-set.

\subsection{Result:}
In  the  proposed  approach  we  use for  model  training \textbf{Catagorial-Cross Entropy} as Loss function and \textbf{Adam} as Optimizer.
\subsubsection{Two-Class Clasification:}

\subsubsection{U-Net:}
Here we train U-Net model with 7 epochs, batch size 4. Each epoch takes 26s. After testing $38$ images are marked as Covid-19 out of total 39 Covid-19 affected images and 81 images are marked as normal out of total 81 normal images. One COVID-19 positive image is marked as normal. So the accuracy is 0.99.17, sensitivity is 0.9744 and specificity 1.0000.

Table 1 presents the performance metrics for the U-Net architecture used in the binary classification of chest X-ray images into COVID-19 and Normal categories. The model demonstrates exceptional performance with a precision of 1.00 for COVID-19 and 0.99 for Normal cases, indicating a high accuracy in predicting positive cases correctly. The recall, which measures the model's ability to identify all actual positive cases, is 0.97 for COVID-19 and 1.00 for Normal cases. The F1 Score, a harmonic mean of precision and recall, is equally high at 0.99 for both classes, showcasing the model's balanced performance in accurately detecting both COVID-19 and Normal cases.

\textbf{The ROC Curve:} Figure 4 illustrates the Receiver Operating Characteristic (ROC) curve for the U-Net model used in the binary classification of chest X-ray images into COVID-19 and Normal categories. The ROC curve plots the true positive rate (sensitivity) against the false positive rate (1-specificity) across different threshold settings, providing a visual representation of the model's performance. The area under the curve (AUC) is 0.98702, indicating that the U-Net model has excellent discriminatory power and is highly effective at distinguishing between COVID-19 positive and Normal cases. A higher AUC close to 1.0 suggests that the model is very accurate, with a minimal chance of misclassification.

\begin{minipage}{\textwidth}
  \begin{minipage}[b]{0.45\textwidth}
    \centering
    \includegraphics[scale=0.65]{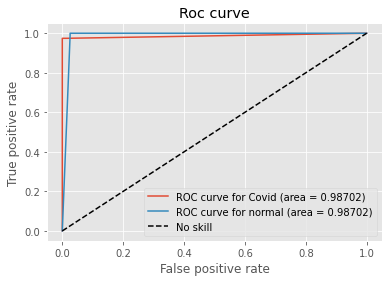}
    \captionof{figure}{ROC curve U-NET }
    \label{fig:myfigure}
  \end{minipage}
  \hfill
  \begin{minipage}[b]{0.45\textwidth}
    \centering
    \includegraphics[scale=0.65]{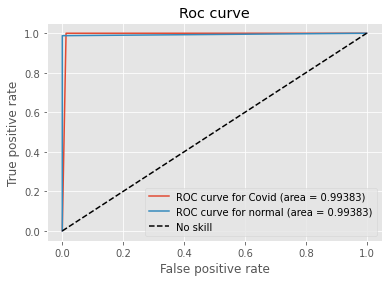}
    \captionof{figure}{ROC curve W-NET }
    \end{minipage}
  \end{minipage}

Table 2 displays the confusion matrix for the U-Net architecture used in the binary classification of chest X-ray images into COVID-19 and Normal categories. The model correctly identified 38 out of 39 COVID-19 positive cases (true positives) and accurately classified all 81 Normal cases (true negatives). There was only one instance where a COVID-19 positive case was incorrectly classified as Normal (false negative), and no Normal cases were misclassified as COVID-19 (false positives). This confusion matrix highlights the U-Net model's high accuracy, with minimal errors in classifying COVID-19 and Normal cases.

\begin{minipage}{\textwidth}
  \begin{minipage}[b]{0.45\textwidth}
    \centering
    \captionof{table}{\label{tab:table1}The performance matrices for U-Net architecture }
\scalebox{1.3}{
\begin{tabular}{ | c | c | c | c | }
\hline
Type & Precision & Recall & F1 Score \\
\hline
Covid & 1.00  & 0.97 & 0.99  \\
\hline
Normal & 0.99 & 1.00 & 0.99 \\
\hline
\end{tabular}}
  \end{minipage}
  \hfill
  \begin{minipage}[b]{0.45\textwidth}
    \captionof{table}{\label{tab:table2}The confusion matrix of U-Net architecture }
\scalebox{0.9}{
\begin{tabular}{ | c | c | c | }
\hline
   & Predicted Positive(Covid) & Predicted Negative\\
\hline
Actual Positive(Covid) & 38  & 1  \\
\hline
Actual Negative & 0 & 81 \\
\hline
\end{tabular}}
    \end{minipage}
  \end{minipage}

Figure 6(a) displays the heat map for the U-Net model used in the binary classification of chest X-ray images into COVID-19 and Normal categories. The heat map visually represents the areas of the X-ray images that the model focused on when making its predictions. Brighter regions in the heat map indicate areas that contributed most significantly to the model's decision-making process. This visualization helps to understand how the U-Net model identifies COVID-19 features in the chest X-ray images and ensures that the model is focusing on relevant areas, such as regions in the lungs, when distinguishing between COVID-19 and Normal cases. The heat map confirms the model's effectiveness by highlighting the critical areas used in its accurate classification.

\subsubsection{W-Net}

We train W-Net model with 10 epochs, batch size 4. Each epoch takes 22s. After testing 39 images are marked as COVID-19 out of total 39 Covid affected images and 80 images are marked as normal out of total 81 normal images. Here one normal image is marked as COVID-19 positive. So the accuracy is 0.9917, sensitivity is 1.000  and specificity is 0.9877.

Table 3 presents the performance metrics for the W-Net architecture in the binary classification of chest X-ray images into COVID-19 and Normal categories. The model exhibits robust performance, with a precision of 0.97 for COVID-19 cases and 1.00 for Normal cases, indicating a high degree of accuracy in correctly identifying positive cases. The recall, which measures the model's ability to detect all actual positive cases, is perfect at 1.00 for COVID-19 and 0.99 for Normal cases, demonstrating the model's effectiveness in identifying true positives. The F1 Score, which balances precision and recall, is also high at 0.99 for both categories, underscoring the W-Net model's reliability in accurately classifying both COVID-19 and Normal cases.

\textbf{The ROC Curve:} Figure 5 illustrates the Receiver Operating Characteristic (ROC) curve for the W-Net model used in the binary classification of chest X-ray images into COVID-19 and Normal categories. The ROC curve plots the true positive rate (sensitivity) against the false positive rate (1-specificity) at various threshold settings, providing a visual assessment of the model's classification performance. The area under the curve (AUC) for the W-Net model is 0.99383, indicating outstanding performance in distinguishing between COVID-19 positive and Normal cases. An AUC value close to 1.0 reflects the model's high accuracy and minimal risk of misclassification, showcasing the W-Net model's superior ability to correctly identify COVID-19 cases compared to Normal cases.

\begin{minipage}{\textwidth}
  \begin{minipage}[b]{0.45\textwidth}
    \centering
    \captionof{table}{\label{tab:table1}The performance matrices for W-Net architecture  }
\scalebox{1.3}{
\begin{tabular}{ | c | c | c | c | }
\hline
Type & Precision & Recall & F1 Score \\
\hline
Covid & 0.97  & 1.00 & 0.99  \\
\hline
Normal & 1.00 & 0.99 & 0.99 \\
\hline
\end{tabular}}
  \end{minipage}
  \hfill
  \begin{minipage}[b]{0.45\textwidth}
    \captionof{table}{\label{tab:table1}The confusion matrix of W-Net architecture }
\scalebox{0.9}{
\begin{tabular}{ | c | c | c | }
\hline
   & Predicted Positive(Covid) & Predicted Negative\\
\hline
Actual Positive(Covid) & 39  & 0  \\
\hline
Actual Negative & 1 &  80 \\
\hline
\end{tabular}}
    \end{minipage}
  \end{minipage}

Table 4 displays the confusion matrix for the W-Net architecture used in the binary classification of chest X-ray images into COVID-19 and Normal categories. The model demonstrates excellent classification performance, correctly identifying all 39 COVID-19 positive cases (true positives) with no false negatives, meaning no COVID-19 cases were missed. It also accurately classified 80 out of 81 Normal cases (true negatives), with only one Normal case being misclassified as COVID-19 (false positive). This confusion matrix highlights the W-Net model's near-perfect accuracy, with a very low rate of misclassification in distinguishing between COVID-19 and Normal cases.

Figure 6(b) displays the heat map for the W-Net model used in the binary classification of chest X-ray images into COVID-19 and Normal categories. The heat map provides a visual representation of the areas within the X-ray images that the W-Net model focused on during its classification process. Brighter regions on the heat map indicate the parts of the image that had the most influence on the model's predictions. This visualization helps to ensure that the model is concentrating on clinically relevant areas, such as the lung regions, when differentiating between COVID-19 and Normal cases. The heat map confirms that the W-Net model accurately identifies the key features associated with COVID-19, thereby validating its high classification performance.

\begin{figure}[h]
\centering
\begin{subfigure}{.49\textwidth}
  \centering
  \includegraphics[width=.8\linewidth]{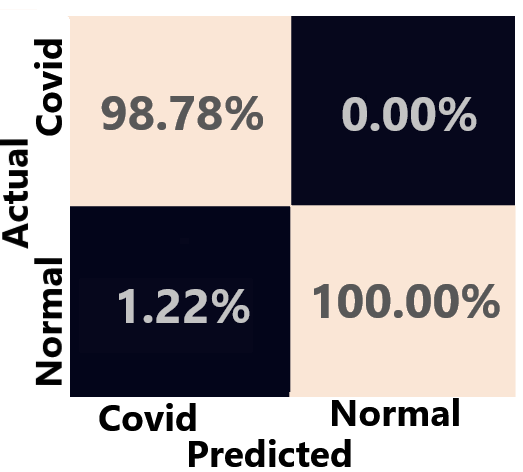}  
  \caption{Heat-map for U-Net}
  \label{fig:sub-first}
\end{subfigure}
\begin{subfigure}{.49\textwidth}
  \centering
  \includegraphics[width=.8\linewidth]{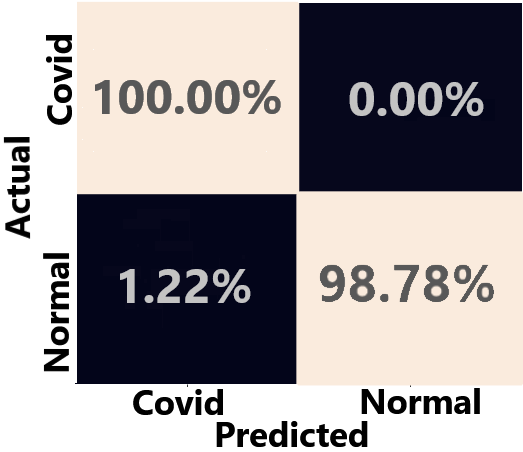}  
  \caption{Heat-map for W-Net}
  \label{fig:sub-second}
\end{subfigure}
\caption{Heat-map}
\label{fig:fig}
\end{figure}

\subsubsection{Three-Class Clasification:}
\subsubsection{U-Net:}
Here we train U-Net model with 13 epochs, batch size 4. Each epoch takes 23s. After testing $40$ images are marked as Covid-19 out of total 40 Covid-19 affected images, 77 images are marked as normal out of total 80 normal images and 77 images are marked as Pneumonia out of 80 Pneumonia affected images. Here one Pneumonia image is marked as Covid-19. So the model accuracy is 0.97, for Covid-19 class - sensitivity is 1.00 and specificity is 0.9935, for Normal class - sensitivity is 0.9625 and specificity is 0.9831 and for Pneumonia class - sensitivity is 0.9625 and specificity is 0.9750.\newline

able 5 presents the performance metrics for the U-Net architecture in the three-class classification of chest X-ray images into COVID-19, Normal, and Pneumonia categories. The model shows strong performance across all classes. For COVID-19, it achieved a precision of 0.98, perfect recall of 1.00, and an F1 score of 0.99, indicating the model’s excellent ability to accurately detect COVID-19 cases. For the Normal class, the precision and F1 score were both 0.97, with a recall of 0.9625, demonstrating the model's reliability in identifying Normal cases. For the Pneumonia class, the model also performed well, with a precision of 0.96, recall of 0.9625, and an F1 score of 0.96, indicating balanced performance in detecting Pneumonia cases. The specificity was notably high for all classes, with the highest being 0.9935 for COVID-19, reflecting the model's ability to accurately distinguish between the different conditions.

Figure 9(a) displays the heat map for the U-Net model used in the three-class classification of chest X-ray images into COVID-19, Normal, and Pneumonia categories. The heat map visually highlights the regions of the X-ray images that the U-Net model focused on during the classification process. Brighter areas on the heat map indicate regions that had the most significant influence on the model's predictions. This visualization helps to understand how the U-Net model identifies distinguishing features for each class, such as specific lung patterns associated with COVID-19, Normal, or Pneumonia cases. The heat map confirms that the model is effectively focusing on relevant areas of the chest X-rays, thereby supporting the model's high accuracy in classifying these three conditions.

\begin{table}[ht]
\centering
\caption{\label{tab:table1}The performance matrices for U-Net architecture}
\scalebox{1.3}{
\begin{tabular}{ | c | c | c | c |c | }
\hline
Type & Specificity &Precision & Recall & F1 Score \\
\hline
Covid & 0.9935  & 0.98 & 1.00 & 0.99  \\
\hline
Normal & 0.9831 & 0.97 & 0.9625& 0.97 \\
\hline
Pneumonia& 0.9750& 0.96& 0.9625&0.96\\
\hline
\end{tabular}}
\end{table}

Table 6 provides the confusion matrix for the U-Net architecture used in the three-class classification of chest X-ray images into COVID-19, Normal, and Pneumonia categories. The model correctly identified all 40 COVID-19 cases (true positives) with no misclassifications (false negatives or false positives). For the Normal class, the model correctly classified 77 out of 80 cases (true negatives), with 3 cases incorrectly classified as Pneumonia (false negatives). In the Pneumonia category, the model correctly identified 77 out of 80 cases (true positives), with 2 cases misclassified as Normal and 1 case incorrectly classified as COVID-19 (false positives). This confusion matrix illustrates the U-Net model's strong performance in accurately classifying COVID-19 cases and its reasonable accuracy in distinguishing between Normal and Pneumonia cases, with only a few misclassifications.

\begin{table}[ht]
\centering
\caption{\label{tab:table1}The confusion matrix of U-Net architecture}
\scalebox{1.3}{
\begin{tabular}{ | c | c | c | c | }
\hline
   & Predicted Covid & Predicted Normal & Predicted Pneumonia\\
\hline
Actual Covid & 40  & 0&0  \\
\hline
Actual Normal & 0 & 77&3 \\
\hline
Actual Pneumonia & 1&2&77\\
\hline
\end{tabular}}
\end{table}

\textbf{The ROC Curve:} Figure 7 illustrates the Receiver Operating Characteristic (ROC) curve for the U-Net model used in the three-class classification of chest X-ray images into COVID-19, Normal, and Pneumonia categories. The ROC curve demonstrates the model's ability to distinguish between these three classes by plotting the true positive rate (sensitivity) against the false positive rate (1-specificity) for each category. The curve provides a visual assessment of the model's performance across different threshold settings. A higher area under the curve (AUC) indicates better classification performance, with the U-Net model showing strong discriminatory power in accurately identifying and differentiating between COVID-19, Normal, and Pneumonia cases. This figure highlights the model's effectiveness in multi-class classification, reflecting its ability to handle complex medical imaging tasks.

\begin{figure}[ht]
\centering
    \includegraphics[scale=0.3]{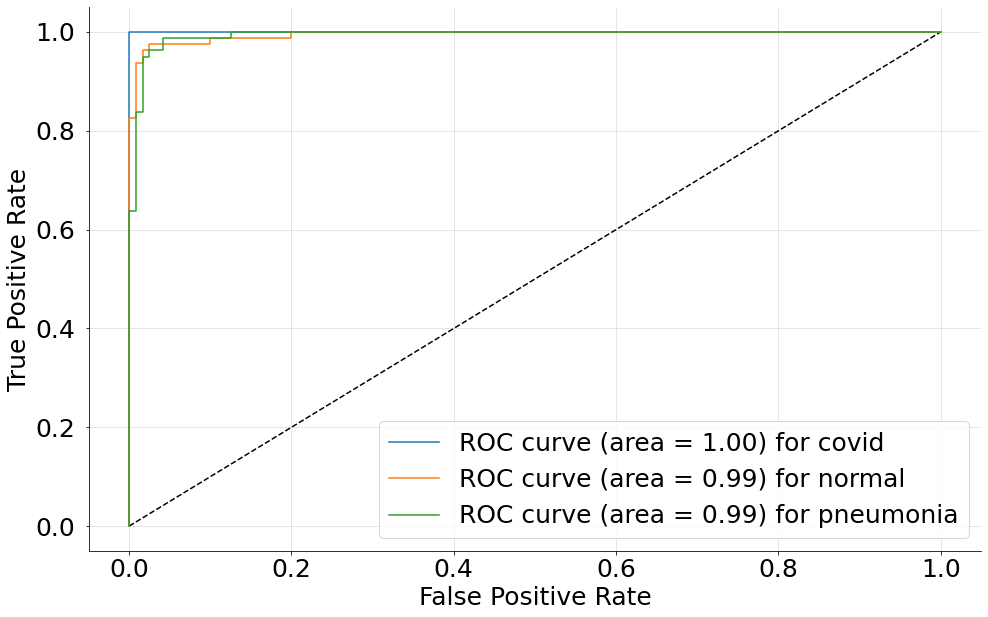}
    \caption{ROC curve U-NET }
    \label{fig:myfigure}
\end{figure}

\subsubsection{W-Net:}
Here we train W-Net model with 12 epochs, batch size 4. Each epoch takes 23s. After testing $40$ images are marked as Covid-19 out of 40 Civid-19 images, 78 images marked as normal out of 80 normal images and 77 images are marked as Pneumonia out of 80 Pneumonia affected images. So the accuracy is 0.9750, for Covid-19 class sensitivity is 1.00 and specificity is 1.00, for Normal class sensitivity is 0.9750 and specificity is 0.9750 and for Pneumonia class sensitivity is 0.9625 and specificity is 0.9833.

Table 7 presents the performance metrics for the W-Net architecture in the three-class classification of chest X-ray images into COVID-19, Normal, and Pneumonia categories. The model shows exceptional performance, with perfect precision, recall, and F1 score of 1.00 for the COVID-19 class, indicating flawless identification of COVID-19 cases. For the Normal class, the model achieved a precision of 0.96, a recall of 0.9750, and an F1 score of 0.97, demonstrating high accuracy in identifying Normal cases with minimal misclassifications. In the Pneumonia category, the model also performed well, with a precision of 0.97, recall of 0.9625, and an F1 score of 0.97, reflecting a balanced and effective classification of Pneumonia cases. The specificity is notably high across all classes, with 1.00 for COVID-19, 0.9750 for Normal, and 0.9833 for Pneumonia, highlighting the W-Net model's strong ability to correctly distinguish between these three conditions.

\begin{table}[ht]
\centering
\caption{\label{tab:table1}The performance matrices for W-Net architecture}
\scalebox{1.3}{
\begin{tabular}{ | c | c | c | c |c | }
\hline
Type & Specificity &Precision & Recall & F1 Score \\
\hline
Covid & 1.0  & 1.0 & 1.00 & 1.0  \\
\hline
Normal & 0.9750 & 0.96 & 0.9750 & 0.97 \\
\hline
Pneumonia& 0.9833& 0.97 & 0.9625 & 0.97\\
\hline
\end{tabular}}
\end{table}

Table 8 displays the confusion matrix for the W-Net architecture in the three-class classification of chest X-ray images into COVID-19, Normal, and Pneumonia categories. The model accurately classified all 40 COVID-19 cases (true positives) without any misclassifications (no false negatives or false positives). For the Normal class, it correctly identified 78 out of 80 cases (true negatives), with 2 cases incorrectly classified as Pneumonia (false negatives). In the Pneumonia category, the model correctly classified 77 out of 80 cases (true positives), with 3 cases misclassified as Normal and none as COVID-19. This confusion matrix highlights the W-Net model's strong performance in accurately detecting COVID-19 cases and its high accuracy in distinguishing between Normal and Pneumonia cases, with only a small number of misclassifications.

\begin{table}[ht]
\centering
\caption{\label{tab:table1}The confusion matrix of W-Net architecture}
\scalebox{1.3}{
\begin{tabular}{ | c | c | c | c | }
\hline
   & Predicted Covid & Predicted Normal & Predicted Pneumonia\\
\hline
Actual Covid & 40  & 0&0  \\
\hline
Actual Normal & 0 & 78&2 \\
\hline
Actual Pneumonia & 0&3&77\\
\hline
\end{tabular}}
\end{table}
\textbf{The ROC Curve:} Figure 8 presents the Receiver Operating Characteristic (ROC) curve for the W-Net model used in the three-class classification of chest X-ray images into COVID-19, Normal, and Pneumonia categories. The ROC curve plots the true positive rate (sensitivity) against the false positive rate (1-specificity) for each class, offering a visual representation of the model's performance across various thresholds. The area under the curve (AUC) for each class indicates how well the model distinguishes between the different categories. A higher AUC value, close to 1.0, reflects the W-Net model's strong ability to accurately differentiate between COVID-19, Normal, and Pneumonia cases. This figure underscores the model's high classification accuracy and its effectiveness in managing multi-class medical imaging tasks, demonstrating superior performance in identifying and separating these three critical conditions.

\begin{figure}[ht]
\centering
    \includegraphics[scale=0.3]{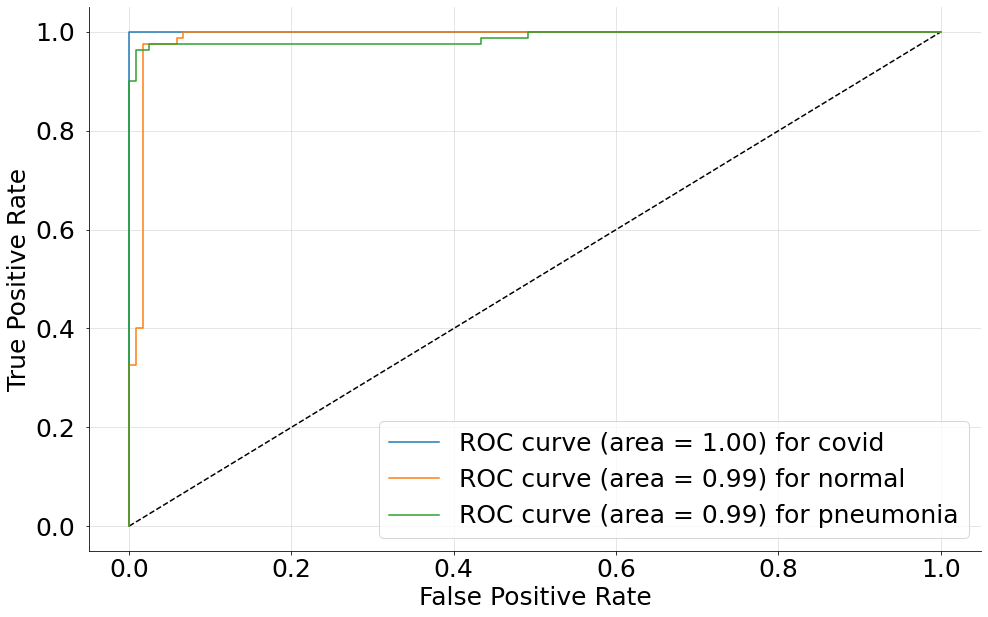}
    \caption{ROC curve W-NET }
    \label{fig:myfigure}
\end{figure}

Figure 9(b) displays the heat map for the W-Net model used in the three-class classification of chest X-ray images into COVID-19, Normal, and Pneumonia categories. This heat map provides a visual representation of the areas within the X-ray images that the W-Net model focused on when making its predictions. Brighter regions in the heat map indicate the most influential areas that guided the model's decision-making process. This visualization is crucial for understanding the model's reasoning, showing that the W-Net model accurately identifies and focuses on the key features in the lungs associated with COVID-19, Normal, and Pneumonia cases. The heat map thus confirms the model's strong performance in distinguishing between these three categories, ensuring that it is concentrating on clinically relevant areas of the images.

\begin{figure}[ht]
\centering
\begin{subfigure}{.495\textwidth}
  \centering
  \includegraphics[width=.9\linewidth]{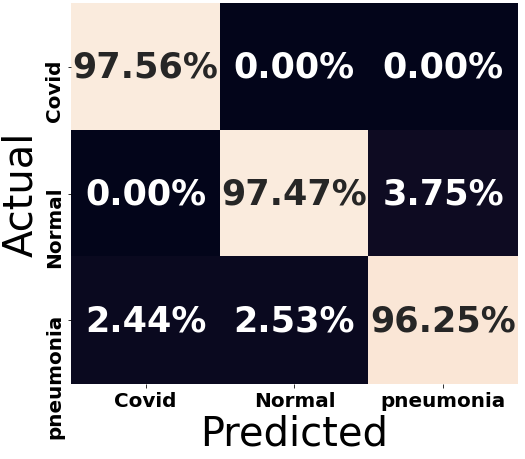}  
  \caption{Heat-map for U-Net}
  \label{fig:sub-first}
\end{subfigure}
\begin{subfigure}{.495\textwidth}
  \centering
  \includegraphics[width=.9\linewidth]{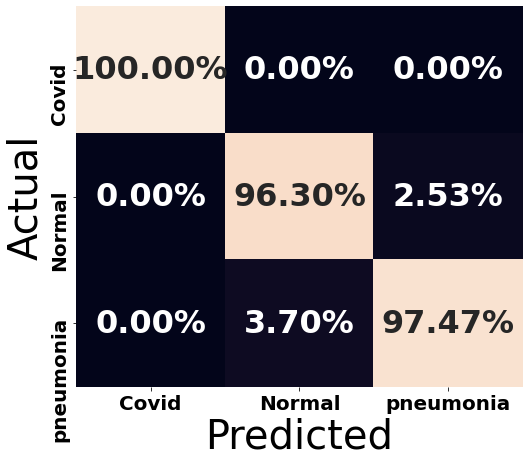}  
  \caption{Heat-map for W-Net}
  \label{fig:sub-second}
\end{subfigure}
\caption{Heat-map}
\label{fig:fig}
\end{figure}

\section{Discussion:}
We perform proposed models performance comparison with existing methods. The comparisons are shown in the Table: 9(for binary classification) and Table: 10(for ternary classification).

\subsection{Binary Classification} Table 9 provides a comparative analysis of various deep learning models used for binary classification (Normal vs. COVID-19) in detecting COVID-19 from chest X-ray images. The models from recent studies demonstrate varying levels of accuracy, with some achieving high sensitivity but not reporting specificity. Mangal et al. (2020) achieved 90.5\% accuracy with 100\% sensitivity, while Narin et al. (2020) reported a 98.08\% accuracy. However, the highest performance was observed in the proposed U-Net and W-Net models from this study, both achieving 99.17\% accuracy. The U-Net model also demonstrated perfect specificity (100\%) with a sensitivity of 97.44\%, while the W-Net model achieved 98.77\% specificity and 100\% sensitivity, outperforming the other models listed in terms of both accuracy and sensitivity.

\begin{table}[ht]
\centering
\caption{Methods comparison for Two class classification (Normal vs. nCovid-19) on Covid 19 detection
}
\scalebox{1.3}{
\begin{tabular}{ | c | c | c | c | }
\hline
Methods & Accuracy & Specificity & Sensitivity \\
\hline
Mangal et.al\cite{a} & 90.5 & - & 100 \\
\hline
Panwar et.al.(2020)\cite{n} & 88.10 & - & -\\
\hline
Hemdan et.al(2020)\cite{o} & 90.00 & - & - \\
\hline
Meghdid et.al(2020)\cite{p}& 94.00 & - & - \\
\hline
Narin et al.\cite{q} & 98.08 &- &- \\
\hline
Azemin,Hassan,Tamrin and Ali(2020)\cite{b} & 71.9 &71.8& 77.3\\
\hline
Ozturk et al. (2020)\cite{e}& 98.08 &- &- \\
\hline
Chandra et. al.(2020)\cite{f}& 98.06 &- &- \\
\hline
Our approach(U-Net based approach)& 99.17 & \textbf{100 }& 97.44 \\
\hline
Our approach(W-Net based approach) & \textbf{99.17} & 98.77 & \textbf{100} \\
\hline
\end{tabular}}
\label{tab:my_label}
\end{table}
\begin{table}[ht]
\centering
\caption{Methods comparison for Three class classification (Normal vs. pneumonia vs nCovid-19) on Covid 19 detection}
    \label{tab:my_label}
\scalebox{1.3}{
\begin{tabular}{ | c | c | c | c | }
\hline
Methods & Accuracy\\
\hline
Ozturk et al. (2020)\cite{e} & 87.02 \\
\hline
L. Wang et al. (2020)\cite{h} & 93.00 \\
\hline
Abbas, Abdelsamea,and Gaber(2020)\cite{i} & 95.12 \\
\hline
Our approach(U-Net based approach)& 97.00   \\
\hline
Our approach(W-Net based approach) & \textbf{97.50} \\
\hline
\end{tabular}}
\end{table}

\subsection{Ternary Classification} Table 10 compares the performance of different deep learning models for three-class classification (Normal vs. Pneumonia vs. COVID-19) in detecting COVID-19 from chest X-ray images. The models exhibit a range of accuracies, with Ozturk et al. (2020) achieving an accuracy of 87.02\%, while Abbas, Abdelsamea, and Gaber (2020) reported a higher accuracy of 95.12\%. The proposed models in this study, based on U-Net and W-Net architectures, showed superior performance with the U-Net model reaching an accuracy of 97.00\% and the W-Net model achieving the highest accuracy of 97.50\%. These results indicate that the proposed models offer more reliable and accurate classification compared to the existing methods.

\section{Conclusion}

In this paper, we present the efficacy of U-net model, W-net model for the prediction of COVID-19 disease from chest X-ray images. Our method exhibits state of the art performance comparing with other modern approaches. Authors are currently engaging to extend this work for multi class classification.

\section*{Acknowledgments}
The authors are really grateful to Dr. Pramit Goswami R.M.O, West Bengal Medical Education Service,  for providing relevant information  and assessment for this work. 

\bibliographystyle{unsrt}  
\bibliography{references}

\providecommand{\noopsort}[1]{}\providecommand{\singleletter}[1]{#1}%
\begin{thebibliography}{10}

\bibitem{a}
Arpan Mangal, Surya Kalia, Harish Rajgopal, Krithika Rangarajan, Vinay Namboodiri, Subhashis Banerjee, and Chetan Arora.
\newblock {\em CovidAID: COVID-19 Detection Using Chest X-Ray}, 2020.

\bibitem{b}
Mohd Zulfaezal, Che Azemin, Radhiana Hassan, Mohd Izzuddin, MohdTamrin, and Md~Ali.
\newblock {\em COVID-19 Deep Learning Prediction Model}, 2020.

\bibitem{c}
Boran Sekeroglu and Ilker Ozsahin.
\newblock {\em Detection of COVID-19 from Chest X-Ray.}, 2020.

\bibitem{d}
Stephanie Stephanie, Thomas Shum, Heather Cleveland, Suryanarayana R.Challa, Allison Herring, Francine~L Jacobson, Hiroto Hatabu, Suzanne CByrne, Kumar Shashi, Tetsuro Araki, Charles S.~White Jose A.~Hernandez, Rydhwana Hossain, Andetta~R Hunsaker, and Mark~M. Hammer.
\newblock {\em Determinantsof Chest X-Ray Sensitivity for COVID- 19}, 2020.

\bibitem{e}
T.~Ozturk, M.~Talo, E.~A. Yildirim, U.~B. Baloglu, O.~Yildirim, and U.~RajendraAcharya.
\newblock {\em Automated detection of COVID-19 cases}, 2020.

\bibitem{f}
Tej~Bahadur Chandra, Kesari Verma, Bikesh~Kumar Singh, Deepak Jain, and Satyabhuwan~Singh Netam.
\newblock {\em Coronavirus disease (COVID-19) detection inChest X-Ray images using majority voting based classifier ensemble}, 165,2021,113909,ISSN 0957-4174, 2001.

\bibitem{g}
Sifat Ahmed, Tonmoy Hossain, Oishee~Bintey Hoque, Sujan Sarker, and Sejuti~Rahman andFaisal Muhammad.
\newblock {\em Automated Covid-19 Detection from Chest X-Ray Images : A High Resolution Network}, 2020.

\bibitem{h}
L.~Wang, Lin, Z.Q., and A~Wong.
\newblock {\em COVID-Net: a tailored deep convolutional neural network design for detection of COVID-19 cases from chest X-ray images}, 2020.

\bibitem{i}
Abbas, A., M.M. Abdelsamea, Gaber, and M.M.
\newblock {\em Classification of COVID-19 in chest X-ray images using DeTraC deep convolutional neural network}, 2021.

\bibitem{j}
J.P. Cohen, Morrison, P., Dao, and L.
\newblock {\em Covid-19 image data collection}, 2020.

\bibitem{k}
Ronneberger, O., Fischer, P., and T~Brox.
\newblock {\em U-net: Convolutional networks for biomedical image segmentation}, 2015.

\bibitem{l}
M.~Abadi et~al.
\newblock {\em TensorFlow: Large-scale machine learning on heterogeneous distributed systems.}, 2016.

\bibitem{m}
Mooney and P.
\newblock {\em \url{ https://www.kaggle.com/paultimothymooney/chest-xray-pneumonia}}, 2018.

\bibitem{n}
Panwar H., Gupta P.K., Siddiqui M.K., Morales-Menendez R., and Singh V.
\newblock {\em Application of deep learning for fast detection of COVID-19 in X-Rays using nCOVnet. Chaos, Solitons and Fractals}, 2020.

\bibitem{o}
Hemdan, E.~E.-D., Shouman, M.~A., Karar, and M.~E.
\newblock {\em COVIDX-Net: A framework of deep learning classifiers to diagnose COVID-19 inX-Ray images}, 2020.

\bibitem{p}
Maghdid, H.~S., Asaad, A.~T., Ghafoor, K.~Z., Sadiq, A.~S., Khan, and M.K.
\newblock {\em Diagnosing COVID-19 pneumonia from X-Ray and CT images using deep learning and transfer learning algorithms}, 2020.

\bibitem{q}
Narin, A., Kaya, C., Pamuk, and Z.
\newblock {\em Automatic detection of coronavirus disease (COVID-19) using X-ray images and deep convolutional neural networks}, 2020.

\end{thebibliography}

\end{document}